\documentclass[12pt]{iopart}
\usepackage{times}
\usepackage{graphicx}

\begin{document}

\title[The crossover from propagating to strongly scattered acoustic modes in d-SiO$_2$]{The crossover from propagating to strongly scattered acoustic modes of glasses observed in densified silica}
\author{E. Courtens, M. Foret, B. Hehlen, B. Ruffl\'{e}, and R. Vacher}

\address{Laboratoire des Verres, UMR 5587 CNRS, Universit\'{e} Montpellier 2, F-34095 Montpellier, France}

\begin{abstract}
Spectroscopic results on low frequency excitations of densified
silica are presented and related to characteristic thermal
properties of glasses.
The end of the longitudinal acoustic branch is marked by a rapid increase of
the Brillouin linewidth with the scattering vector.
This rapid growth saturates at a crossover frequency $\Omega_{\rm co}$
which nearly coincides with the center of the boson peak.
The latter is clearly due to additional optic-like excitations related
to nearly rigid SiO$_4$ librations as indicated by hyper-Raman scattering.
Whether the onset of strong scattering is best described by hybridization
of acoustic modes with these librations, by their elastic scattering
(Rayleigh scattering) on the local excitations, or by soft potentials
remains to be settled.
\end{abstract}

\pacs{63.50+x, 78.35+c, 61.43-j, 78.70.Ck}
\vspace{1cm}

\section{Introduction}

Studying some papers in these proceedings, or judging from recent
review articles \cite{Cou01,Ruo01}, the reader will find that rather
different views regarding the fate of high frequency
acoustic-like modes in glasses are still being presented.
Essentially, the controversy has to do with the length scale
beyond which the continuous homogeneous medium approximation
breaks down in glasses.
One group essentially claims that plane acoustic waves
{\em propagate} with wavelengths down to the near atomic
scale in a large variety of glasses \cite{Ruo01}.
The other group, to which we belong, believes that plane
waves cannot be reasonable approximations to eigenmodes
for wave vectors beyond a crossover value $q _{\rm co}$ \cite{Cou01}.
This $q _{\rm co}$ is such that
$ 2 \pi / q _{\rm co}$
is considerably larger than the structural
units of usual glasses, for example than the SiO$_4$ tetrahedra
in vitreous silica, $v$-SiO$_2$.
As shown below, this can be demonstrated in the case of densified silica.
However, on the basis of {\em all} the available information, it
seems to be a reasonable hypothesis for many glasses.
This is not a side issue but rather a central question for
anyone with interest in the long range structure of
these important materials.
It is also crucial for the understanding of macroscopic
properties that reflect the disorder.
For example, it is well known that the low-temperature thermal
conductivity $\kappa(T)$ of insulators is controlled by the mean
free path $\ell$ of propagating acoustic waves \cite {Kit49,Zel71}.
As the temperature $T$ is raised from very small values, say 0.1 K,
waves of higher frequency $\Omega$ participate to the conductivity
with a nearly constant $\ell$, so that $\kappa$ increases.
However, the thermal conductivity of glasses usually shows a well
pronounced plateau for $T \sim 10$ K, corresponding to dominant
phonons around 1 THz, {\em e.g.} \cite{Zel71,Gra86,Ray89} or also
a chapter in \cite{Phi81}.
The traditional interpretation of this feature is that the plateau
corresponds to an upper limit $\Omega _{\rm co}$ for the frequency
$\Omega$ of {\em propagating} plane waves.
In other words, the plateau corresponds to a crossover where
$\ell \rightarrow 0$ for the
dominant phonons at the plateau temperature.

As the question is of importance, it is natural to turn
to spectroscopy to shed light onto it.
Spectroscopy has established that in strong glasses, like
$v$-SiO$_2$, acoustic waves propagate energy up to rather high
frequencies.
This has been demonstrated with pulse experiments up to
$\simeq 0.3$ THz \cite{Zhu91}.
Low temperature tunneling experiments also revealed in $v$-SiO$_2$ a
linear dispersion, $\Omega = v q$, and narrow linewidths $\Gamma$ up
to at least $\Omega / 2 \pi \simeq 0.4$ THz \cite{Rot84}.
Here, $v$ is the velocity and $q$ is the wave vector of
quasi-plane waves.
For such waves, the spectrum has a half width $\Gamma$
which in cps is $v / 2 \ell$, where 
$\ell$ is the {\em energy} mean free path.
For quasi-plane waves, $\Gamma$ is {\em much} smaller than $\Omega$.
The recent development of high-resolution inelastic x-ray
spectroscopy (IXS) in principle allows to observe the spectrum of
high-frequency acoustic waves by x-ray Brillouin scattering
\cite{Ruo01}.
Here the experimental possibilities are however limited to
sufficiently high spectral frequencies $\omega$ and
scattering vectors $Q$.
This is presumably why the controversy arose in the first place
\cite{For96,Ben96,For97}.
Indeed, many publications claimed that sound {\em propagates} at
frequencies much above the earlier expectations for
$\Omega_{\rm co}$, \cite{Ben96,Mas96,Mas97,Mas98,Fio99,Ruo99,Mas00,Pil00}
this without being really able to perform the
decisive spectroscopy around $\Omega_{\rm co}$ as it is located
too low in $\omega$ to be accessible.
If such claims would be valid, one would have a very hard
time to explain the observations of well defined $\kappa (T)$ plateaus in
glasses \cite{Zel71,Gra86,Ran88}.

In our view, the above claims result from misinterpretations of
the spectroscopic data.
This will be explained for one particular glass below.
Thus, in our opinion the real issue is not whether the crossover frequency
$\Omega_{\rm co}$ exists, which it probably does in many cases,
but rather what is the mechanism that produces it.
Early work assumed that Rayleigh scattering by disorder (whether
disorder in masses or in force constants) would be sufficient to
produce a crossover \cite{Zel71,Gra86}.
Rayleigh scattering by point defects leads to the inhomogeneous
broadening of plane waves with $\Gamma = A \Omega ^4$
\cite{Car61,Kle58,Kle55,Pom42,Mar66}.
With such a high power of $\Omega$, as soon as this broadening
becomes observable, the limit $\Gamma \sim \Omega$ should
be quickly reached as $\Omega$ further increases.
More precisely, the upper limit for plane-wave-like propagation
should rather be $\ell \simeq \lambda /2$ which amounts to
$\Gamma \simeq \Omega / 2\pi$.
This corresponds to the Ioffe-Regel crossover beyond which
$\Gamma \propto \Omega$ \cite{Iof60} and one enters the strong
scattering regime for plane waves.
The limit should essentially coincide with $\Omega_{\rm co}$.
However, the prefactor $A$ appears to be too small by about an
order of magnitude to lead to the expected low value for
$\Omega_{\rm co}$ \cite{Gra86,Ran88,Jon78}, although
\cite{Ray89} claims that force-constant fluctuations are
sufficient in $v$-SiO$_2$ to account for the plateau.
Thus one might wish to search for other possible mechanisms.
A likely one can result from the near universal presence
of another spectral feature of glasses, the
so called ``boson peak'' \cite{Ran88,Dre68,Yu87,Gra90}.
This peak corresponds to additional
excitations which produce a hump in $C/T^3$ {\em vs.} $T$, where
$C$ is the specific heat \cite{Zel71}.
Like the $\kappa$ plateau, this hump is located around 10 K.
It is produced by an excess over the Debye value in the density of
vibrational states (DOS), $Z(\omega )$ \cite{Buc86}.
This excess is well observed by plotting $Z(\omega )/\omega ^2$, a quantity
that should be constant in the Debye model but which also
shows a hump around 1 THz in $v$-SiO$_2$ and in many other glasses.
The question arose whether these boson-peak (BP) excitations
just correspond to the near horizontal end of acoustic
branches in the region where $q$ reaches Brillouin-zone-like dimensions
\cite{Mas99,Fon99},
or whether they rather are due to other modes \cite{Buc84},
which then must be optic-like ones.
A very strong case for the latter has recently been made for
various forms of silica on the basis of hyper-Raman scattering
evidence \cite{Heh00}. This will be briefly explained below. 
It is also known since a long time that in some model systems that
behave as glasses, such as the mixed crystals of KCN-KBr
\cite{Ran88,Gra90}, the boson peak is optic-like in nature,
corresponding in the latter case to CN librations.
In the case of silica, the BP forms a broad diffuse band
that indicates strong inhomogeneous broadening.
The acoustic modes can hybridize with these low-lying optic-like
modes \cite{Kar85,Kli01}.
This hybridization could then produce the crossover, in which case one
should expect that $\Omega_{\rm co}$ might be close to the frequency
$\Omega_{\rm BP}$ of the BP-maximum.\\

\section{Spectroscopy of densified silica glass}

To progress on this issue, it seems that a good way is to perform high
quality Brillouin scattering spectroscopy on well selected cases.
At present, only IXS allows to investigate the
waves at both the relevant frequencies and scattering vectors.
To settle this question, it is necessary to clearly observe $\Omega_{\rm co}$
and the corresponding $q_{\rm co}$, the evolution of the acoustic profile,
as well as the onset of the boson-peak scattering of x-rays.
In this respect, it must be emphasized that IXS
still is a difficult spectroscopy with severe limitations on
resolution and intensity.
Firstly, it is practically not possible with current instruments to
investigate scattering vectors $Q$ below $\sim 1$ nm$^{-1}$.
Secondly, the narrowest instrumental profile allowing for sufficient intensity
\cite{Ver96} still has an energy
width around 1.5 meV (or $\simeq 0.4$ THz) and extended Lorentzian-like wings.
Owing to the relatively strong elastic scattering of glasses,
this tends to mask the weaker Brillouin signal.
Finally, the signal-to-noise ratio being also small, the subtle changes in
Brillouin lineshapes that indicate strong scattering might easily
go unnoticed \cite{For97}.
For the above reasons, the necessary spectroscopy cannot yet be
performed on normal vitreous silica, $v$-SiO$_2$.
In that case, the position of the $\kappa$ plateau around 10 K suggests that
$\Omega_{\rm co} \simeq 4$ meV (or $\Omega_{\rm co} / 2 \pi
\simeq 1$ THz) so that one expects $q_{\rm co} \simeq 1$ nm$^{-1}$
for the longitudinal acoustic waves of velocity $v \simeq 5900$ m/s.
These values of $\Omega_{\rm co}$ and $q_{\rm co}$ are indeed too low
to clearly observe crossover phenomena in $v$-SiO$_2$.
To alleviate several of these difficulties, we investigated
another form of silica glass, permanently densified silica $d$-SiO$_2$,
in which case we found that $\Omega_{\rm co}/2\pi \simeq 2$ THz and
$q_{\rm co} \simeq 2$ nm$^{-1}$ \cite{Rat99,For02}.

That silica can be permanently densified when subjected to high pressures
has been known for a long time \cite{Bri53}.
The sample of $d$-SiO$_2$ used in all our IXS measurements was densified
at $\sim 1000$ K and 7.4 GPa \cite{Ina98}.
It was kindly provided by Dr. M. Arai.
It is a clear, transparent piece, whose good homogeneity has been
checked by optical Brillouin spectroscopy \cite{RatTh99}.
Its density is $\rho \approx 2.60$ g/cm$^3$, much above that of $v$-SiO$_2$,
$\rho = 2.20$ g/cm$^3$, and near that of crystal quartz.
Intuitively, one expects that the range of inhomogeneities should be
strongly reduced by densification, and hence that $q_{\rm co}$ and
$\Omega_{\rm co}$ should be larger in $d$-SiO$_2$ than in $v$-SiO$_2$.
An increase of $\Omega_{\rm co}$ is indeed indicated by the $\kappa (T)$
data.
The plateau seems to be around 20 K in $d$-SiO$_2$ \cite{DMZ94}, suggesting
that $\Omega_{\rm co}$ will be roughly twice that of $v$-SiO$_2$.
Also, the disturbing signal from the elastic structure factor, $S(Q,0)$,
should be reduced considerably compared to that of $v$-SiO$_2$ \cite{Ina98}.
For these reasons we anticipated that the crucial region below and near
$\Omega_{\rm co}$ might be accessible to spectroscopy in $d$-SiO$_2$
using the current IXS capabilities.\\

\begin{figure}
\centering\includegraphics[width=12cm]{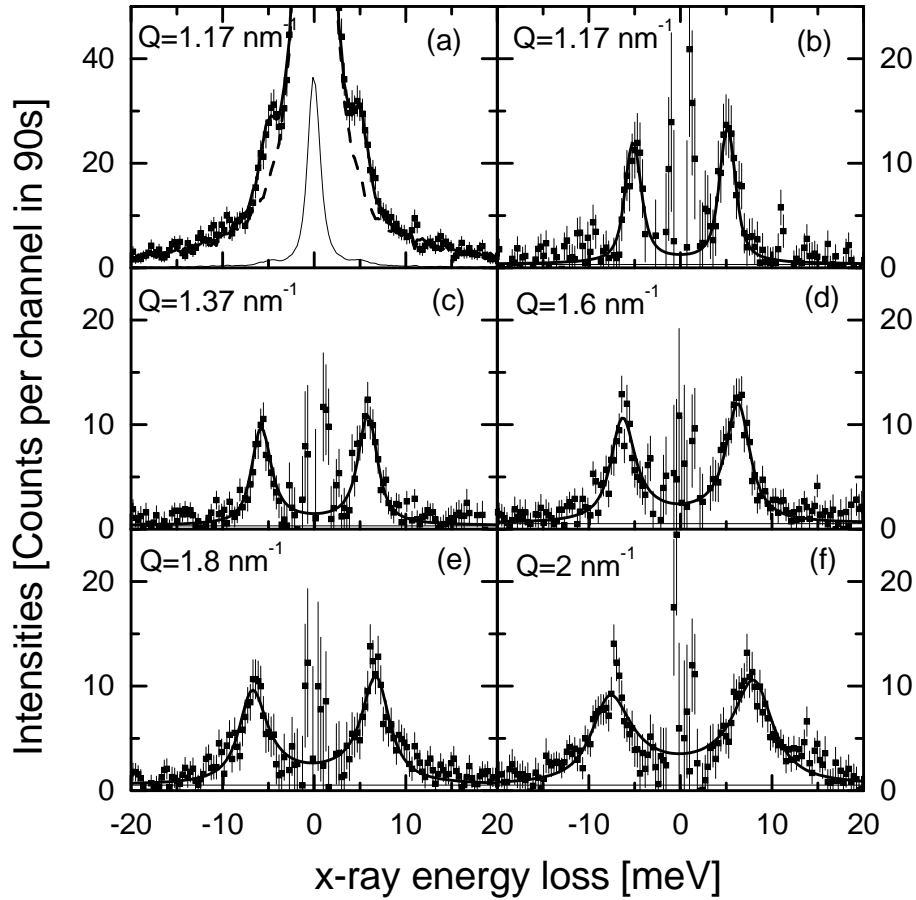}
\caption{IXS spectra of $d$-SiO$_2$.
(a) A full spectrum with its central part, divided by 20, shown
by the solid line.
The dashed line shows the adjusted elastic contribution.
(b) to (f) The inelastic part for five different scattering vectors
with the solid lines showing the DHO fits.
The fitted elastic component was subtracted from the data (see text).
The level of the electronic dark counts is shown by the solid lines.}
\end{figure}

\section{The crossover to strong scattering}

Figure 1 shows a series of IXS spectra of $d$-SiO$_2$ taken at small
values of $Q$, up to the expected $q_{\rm co}$ \cite{Ruf03}.
These spectra were obtained on the spectrometer ID16 at the
European Synchrotron Radiation Facility in Grenoble, France.
The experimental conditions are described in \cite{Rat99,Ruf03}. 
As shown in figure 1(a), the spectra are dominated by an elastic central peak.
To increase the relative strength of the inelastic contribution, the sample is
placed at an elevated temperature which increases the thermal
population of the acoustic modes.
However, $d$-SiO$_2$ relaxes towards $v$-SiO$_2$ if $T$ is too high.
We found that a good compromise is $T = 565$ K, as at that value the
density does not change over periods of weeks \cite{RatTh99}.
To obtain significant information on the inelastic Brillouin doublet,
it is necessary to have an instrumental lineshape which is extremely
clean far into its wings.
This has been the case in this particular experiment \cite{Ruf03}.
The instrumental lineshape was determined for each analyzer 
using the signal of polymethyl methacrylate at 20 K.
Very clean Voigt-like profiles were obtained with half-widths at
half-maximum of $\simeq 0.75$ meV.
The spectra can thus be fitted to the sum of an elastic line of integrated
intensity $I_{\rm CP}$ plus a suitable spectral function $S(Q,\omega)$.
Below the crossover and in absence of any other information, it is
reasonable to select for $S(Q,\omega)$
a standard damped harmonic oscillator (DHO) response function.
In this way one obtains a frequency $\Omega$, a half width $\Gamma$
and an integrated intensity $I_{\rm DHO}$ independently from any other
assumption.
Of course the linewidth will be a combination of a homogeneous
(or lifetime) broadening $\Gamma_{\rm hom}$ and of any inhomogeneous broadening
$\Gamma_{\rm inh}$ that might result from the approach of the crossover.
To evaluate the spectra,
the response function must be convoluted with the instrumental
lineshape.
To extract significant inelastic widths it is necessary to take into
account the collection angles that are fixed by the slits placed in front
of the analysers.
These give an uncertainty in the scattering vectors, $\Delta Q = \pm 0.18$
nm$^{-1}$.
The fitting procedure sums over surface elements of the slit, each at its
own $Q$, using $d\Omega /dQ = v_{\rm g}$, where $v_{\rm g}$ is the group
velocity which is determined iteratively.
Figures 1(b) to 1(f) represent the pure inelastic part that remains
after subtraction of the central peak of adjusted strength $I_{\rm CP}$.
The solid lines are the convoluted DHO fits.
The half-width $\Gamma$ obtained for $Q=2$ nm$^{-1}$ is somewhat larger than
the corresponding $\Omega /2 \pi$.
Hence, that value of $Q$ is already in the crossover region \cite{Rat99}.
This confirms the estimate made on the basis of the $\kappa$
plateau \cite{DMZ94} and it shows that in $d$-SiO$_2$
the crossover region is accessible.

\begin{figure}
\centering\includegraphics[width=12cm]{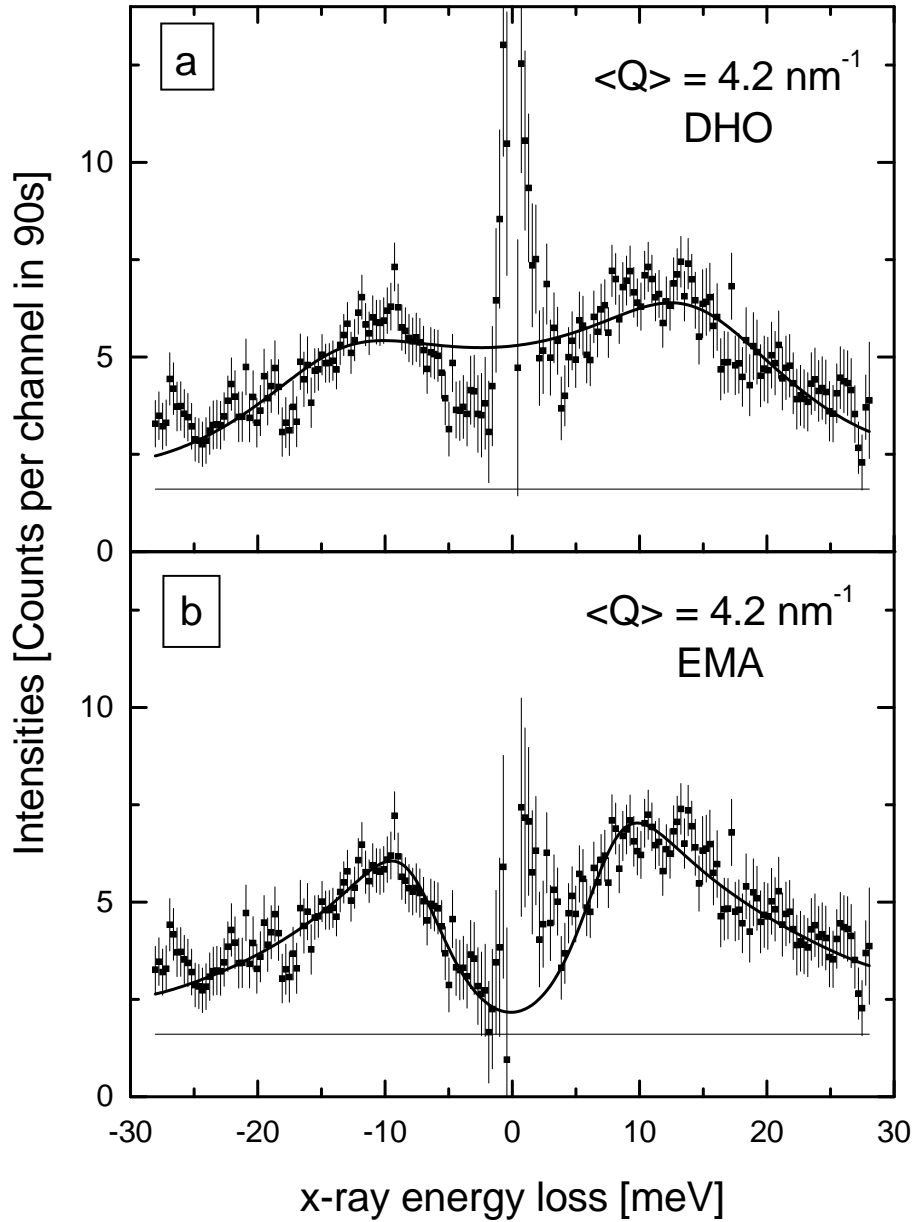}
\caption{The inelastic part of IXS spectra of $d$-SiO$_2$ for
$Q > q_{\rm co}$.
The solid lines are fits to the DHO (a) and EMA (b) spectral profiles.
The dark-count baselines are shown.}
\end{figure}

Figure 2 shows a spectrum obtained for $Q > q_{\rm co}$.
Beyond the crossover the spectra evolve quite slowly with $Q$.
What is shown in figure 2 is the sum of spectra obtained at $Q= 4$, 4.2, and
4.4 nm$^{-1}$ in the course of accumulating the spectra shown in figure 1.
In this manner the total accumulation time in figure 2 is nearly one week.
This summation improves the signal-to-noise ratio without
changing appreciably the spectral shape.
For  $Q > q_{\rm co}$, one does not expect that the DHO can be a
reasonable approximation and
indeed, the DHO fit in figure 2(a) is not satisfactory.
In particular, it gives a contribution near $\omega = 0$ which is
much higher than what the data suggest, this in spite of the freely
adjustable  $I_{\rm CP}$ which tries to compensate for it.
The reason is that for  $q > q_{\rm co}$ there are no eigenmodes
with a well defined $q$-value.
For a given $\Omega > \Omega_{\rm co}$, the eigenmodes consist in a broad
superposition of plane waves in $q$-space.
The measurement being performed at a fixed scattering vector $Q$, the
spectrum is the projection on $Q$ of the appropriate Fourier components
for all $\Omega$'s  \cite{Vac99}.
On the other hand,
at low frequencies $\omega$, the modes with $\Omega = \omega$ are well
defined plane waves of a given $q << q_{\rm co}$.
These do not have Fourier components at $Q >> q_{\rm co}$ and
for this reason the spectrum dips in its center.
To approximately describe this situation, one can take an EMA
(Effective Medium Approximation) in which the inhomogeneity is described by
frequency dependent velocity, $v(\omega)$, and linewidth, $\Gamma ( \omega )$.
Figure 2(b) shows the result of such an EMA adjustment as described in
\cite{Rat99}.
One should note that once the parameters of the EMA are fixed,
they allow predicting both
the spectral shape and strength at all $Q$-values.

\begin{figure}
\centering\includegraphics[width=12cm]{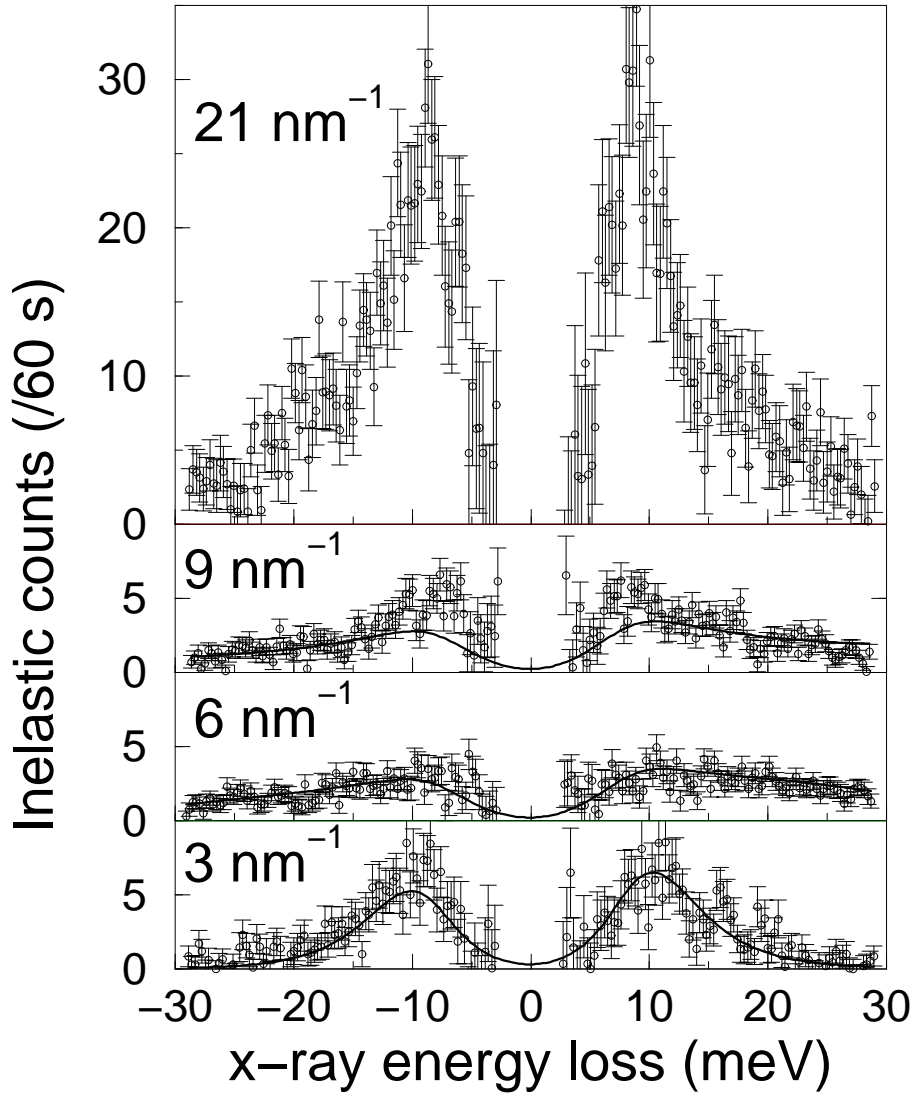}
\caption{Inelastic part of IXS spectra at scattering vectors above
the crossover.
The solid lines show the prediction of the EMA models with the
parameters obtained from the fits in reference \cite{Rat99}.
The dark counts have been subtracted.}
\end{figure}

Figure 3 shows the evolution of the IXS spectra deeper into the strong
scattering regime, {\em i.e.} for $Q$ growing beyond $q_{co}$ \cite{For02}.
The spectra are shown on a single relative scale, uncorrected for the atomic
form factors.
The form factors reduce the observed intensity by 14 \% at 9
nm$^{-1}$ and by nearly a factor of two at 21 nm$^{-1}$.
The solid lines are the predictions of the same EMA as in figure 2,
taking into account the effect of the atomic form factors.
At 9 nm$^{-1}$, an extra signal appears on top of the EMA which then
grows rapidly to become overwhelming at 21 nm$^{-1}$.
This is the scattering from the BP excitations.
These BP spectra are similar to those that have been observed with inelastic
neutron scattering \cite{Buc86}.
They have, in that case, been discussed within an ``incoherent
approximation'' \cite{Buc85}.
This essentially amounts to assuming that $2 \pi /Q$ is small compared
to the coherence length of the vibrating modes so that on the average one
observes single vibrating units that can be thought as incoherently moving
with respect to one another.
In that case, the quantity which is observed is essentially
$ \propto n(\omega) Z(\omega)/\omega$, where $n(\omega)$ is the Bose factor.
This is easily converted to  $Z(\omega)/\omega ^2$,
as displayed {\em e.g.} in \cite{Ina99}.
The rapid increase of intensity in figure 3 is consistent with an
approximate growth with $Q^2$ \cite{Buc86}.
The main point is that the BP maximum, $\Omega_{BP}$, is seen to practically
coincide with $\Omega_{co}$.
It is thus of considerable interest to investigate the origin of these
BP modes.\\

\section{The origin of the boson peak of silica glasses}

The nature of the BP in glasses has long been discussed.
Qualitatively similar BP spectra are often observed in various
spectroscopies at
scattering vectors $Q$ which differ by orders of magnitude.
This suggests that these excitations are quasi-local.
However, a difficulty arose in silica from the fact that the inelastic
neutron scattering (INS) spectra and the Raman scattering (RS) ones are
quantitatively quite different \cite{Fon99}.
If {\em the same} modes are observed in both spectroscopies, one
would expect that the RS intensity is $\propto Z(\omega)
n(\omega) C(\omega)/\omega$, where $Z(\omega)$ is the DOS
observed with neutrons \cite{Shu70}.
The coupling coefficient $C(\omega)$ should be $\propto \omega^2$
for active acoustic modes \cite{Mar74} and should be constant
for active optic ones \cite{Shu70}.
Instead of that, the experimental determination gives
approximately $C(\omega) \propto \omega^1$ in $v$-SiO$_2$ \cite{Fon99}.
To clarify this situation we performed hyper-Raman scattering (HRS)
from various forms of silica glasses, in particular $v$-SiO$_2$
and $d$-SiO$_2$ \cite{Heh00}.
HRS is a nonlinear scattering spectroscopy in which two incoming
photons give one scattered photon \cite{Den87}.
It obeys selection rules which are different from those of RS and
infrared absorption (IR).
In particular, if the effective symmetry is that of
isotropic media, $\infty \infty m$, then there is mutual exclusion
between RS and HRS activities.
This means that acoustic modes, which are active in Brillouin
scattering, are {\em not} active in HRS.
On the other hand, one can consider the vibrational modes of local units of
lower symmetry, like for example the SiO$_4$ tetrahedra.
The rigid rotations of undistorted tetrahedra,
which have a spherically symmetric polarisability,
are not active in RS but they happen to be active in HRS \cite{Cyv65}.

\begin{figure}
\centering\includegraphics[width=12cm]{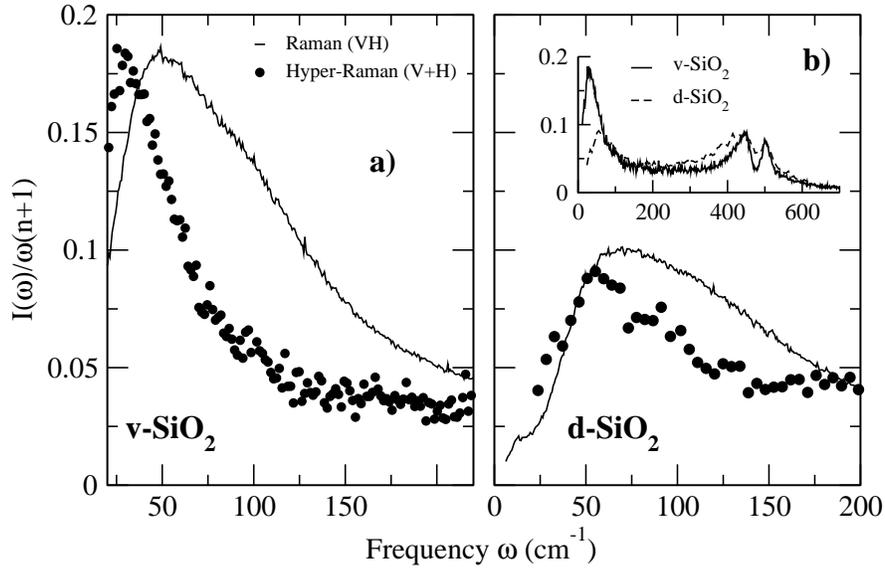}
\caption{Raman and hyper-Raman scattering spectra in $v$-SiO$_2$ (a)
and $d$-SiO$_2$ (b) in the boson peak region.
The inset in (b) shows hyper-Raman scattering spectra in $v$-SiO$_2$
and $d$-SiO$_2$ on a frequency range which allows the observation of
the first TO and LO modes.
The data are drawn on comparable relative scales.}
\end{figure}

Some of the HRS results are illustrated in figure 4.
One sees immediately that the RS and HRS spectra are very different.
In addition, although the HRS
signal is very small, the HRS scattering
{\em activity} is much larger than the RS  one.
The quantitative comparison is explained in \cite{Heh00} where it is
shown that the BP modes are mostly RS inactive while
they are HRS active.
This is a first indication that the BP modes are optic-like.
One also notes the quasi-absence of a BP in IR \cite{Heh00}, another
selection rule which shows that the BP is essentially non-polar.
Comparing with $Z(\omega)/\omega^2$ obtained in INS, one finds a
nearly perfect overlap with the HRS spectra \cite{Heh00}.
This obviously indicates that for HRS the coupling constant $C(\omega)$
is $\propto \omega^0$, which is a second proof that the relevant modes
are optic-like.
On the basis of INS it was already noted that rigid
SiO$_4$ tetrahedra librations are likely to be at the origin of the
BP excitations \cite{Buc86}.
These authors remarked that the $Q$-dependence of the scattered spectra
could be qualitatively reproduced by a model taking into account the
coupled rotations of five rigid SiO$_4$ tetrahedra.
In this respect it is interesting to remark that of all the modes of
a tetrahedral ``molecule'', the only ones which have the required
selection rules, namely to be inactive in RS and IR, and active in HRS,
are the rigid rotations \cite{Cyv65}.

Interestingly, nearly rigid tetrahedra rotations are at the origin of the
$\beta \rightarrow \alpha$ transition in crystal quartz
which occurs around 570 $^\circ$C.
The soft mode of this transition is silent in RS but active in
HRS \cite{Tez91}.
Moreover, the extrapolation of that HRS measurement shows that at
$T \simeq$ 1100 $^\circ$C, the frequency of the soft mode
is approximately located at the position of the BP of $v$-SiO$_2$.
That value of $T$ corresponds to the T$_g$ of $v$-SiO$_2$.
Therefore the BP in the glass can be viewed as a frozen soft mode.
As the freezing is taking place far from the actual transition, the
correlation length of the mode is expected to be relatively short.
In $d$-SiO$_2$, the densification from 2.2 to
2.60 g/cm$^3$ obviously should arrest the collective rotations
which have the largest spatial extent.
The low frequency part of the BP of $v$-SiO$_2$ just disappears
in going to $d$-SiO$_2$.
This is well seen in figure 4(b), in particular in the inset which shows
spectra of $v$-SiO$_2$ and $d$-SiO$_2$ on the same relative scale.
This suggests that the collective excitations with the largest spatial
extent also have the lowest frequencies, just as for a soft mode
near a structural transition.
This inset of figure 4 also shows the lowest TO-LO modes.
The TO develops a broad low-$\omega$ wing in $d$-SiO$_2$
and this wing apparently contributes to the HRS signal in figure 4(b)
at frequencies above $\sim$ 150 cm$^{-1}$.

In view of all the above, it is now clear that the BP of silica is an
optic-like mode related to the nearly rigid rotation of the SiO$_4$ units.
This is also confirmed by several independent recent simulations
\cite{Tar97,Gui97,Pas02}.
This BP is active in HRS, while what is observed in RS is most probably
a leakage of this forbidden mode, either that the tetrahedra are
slightly distorted and therefore that their polarizability is not
fully spherically symmetric, or that the scattering results from
the not fully symmetric near environment of the tetrahedra.
We believe that the unusual exponent found in RS for the
coupling coefficient, $C(\omega) \propto \omega ^ 1$,
precisely reflects the average strength of this leakage of
forbidden modes.
As the BP does not appreciably change its spectral shape with $T$, it is
obviously very strongly inhomogeneously broadened, meaning that
single BP components can be very long-lived quasi-local vibrations
that are distributed randomly in the volume of the glass.
On the other hand, it is intuitive that the SiO$_4$ rotations will
have a bilinear coupling with elastic deformations, and therefore
with acoustic-like modes.\footnote{In $\beta$-quartz, there is a
bilinear coupling between the {\em gradient} of the soft-mode
and the acoustic modes \cite{Asl79}.
That coupling produces the incommensurate phase that appears
between the $\beta$ and $\alpha$ ones \cite{Dol92}.
In the glass, the gradient becomes superfluous owing to the randomness.}
As acoustic modes become resonant with BP modes, this bilinear coupling
produces hybridization which amounts to their strong scattering
owing to the random spatial distribution of the BP modes.
Evidence for this can be searched in the onset of strong scattering
as $\Omega$ increases towards $\Omega_{BP}$.\\

\begin{figure}
\centering\includegraphics[width=12cm]{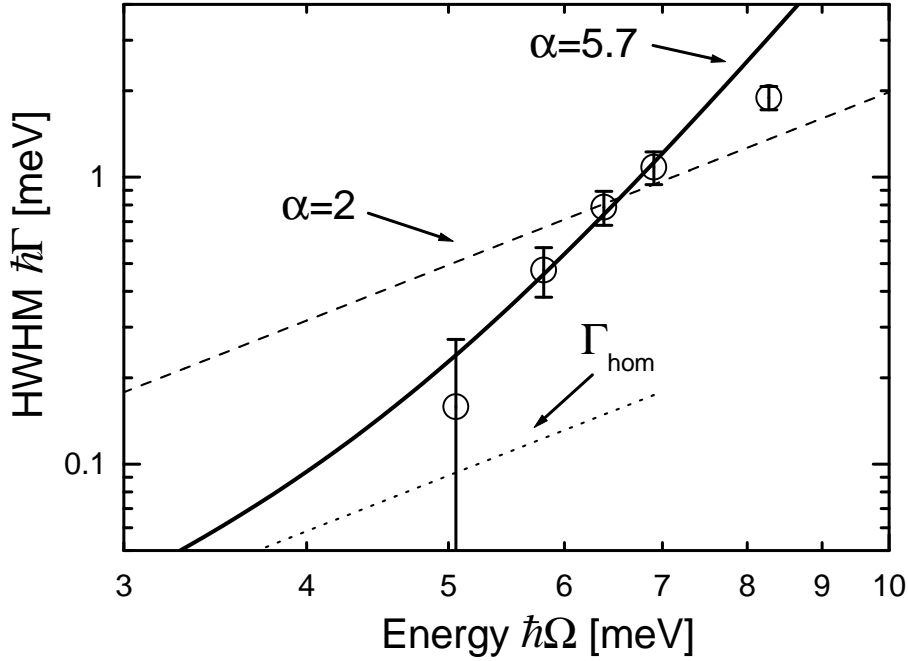}
\caption{The half-widths at half-maximum of the IXS spectra obtained from
the DHO fits shown in Fig. 1. They are presented in function of $\Omega$.
The estimated homogeneous broadening is shown by the dotted line labelled
$\Gamma_{\rm hom}$.
The other lines are fits as described in the text.}
\end{figure}

\section{The onset of strong scattering in densified silica}

The linewidths $\Gamma$ extracted from the DHO fits shown in figure 1
are plotted in figure 5.
One notices a very rapid increase of $\Gamma$ with $\Omega$.
We use Matthiessen's rule that $\Gamma = \Gamma_{\rm hom} +
\Gamma_{\rm inh}$.
To evaluate $\Gamma_{\rm inh}$ we need an estimate for $\Gamma_{\rm hom}$.
Usually the linewidth observed in optical Brillouin scattering is
purely homogeneous, baring macroscopic inhomogeneities in the density
which were carefully avoided \cite{RatTh99}.
We found that in near backscattering, for
$\Omega$ = 41.5 GHz, the $\Gamma$ of $d$-SiO$_2$ is similar to
that of $\alpha$-quartz, and this over a broad $T$-range, from
100 to 300 K \cite{RatTh99}.
As the homogeneous attenuation in $\alpha$-quartz is
of the Akhieser type \cite{Akh39,Boe60}, it is reasonable to assume that
also in $d$-SiO$_2$ one would have approximately
$\Gamma_{\rm hom} \propto \Omega^2$.
This gives the line marked $\Gamma_{\rm hom}$ traced in figure 5.
It is extrapolated from the optical Brillouin value 
$\Gamma_{\rm hom} = 26 \pm 5$ MHz at 41.5 GHz. 
If there would be appreciable relaxation contributions,
as described for $v$-SiO$_2$ in \cite{Wie00}, our extrapolation is
certainly an upper bound.

The measured widths are clearly above the line
$\Gamma_{\rm hom}$ and they increase with $\Omega$
faster than $\Omega^2$.
This can be shown ad absurdum by drawing the best fit
in $\Gamma \propto \Omega^2$.
One obtains the dashed straight line marked $\alpha$ = 2 in figure 5.
The fit gives a mean-square deviation $\chi^2$ = 5.4, which is
very poor indeed.
On the other hand one might attempt a power law fit
$\Gamma_{\rm inh} = A \Omega ^{\alpha}$.
In that case one should remark that 
the point measured at $\Omega > 8$ meV is already
in the crossover region, as discussed above.
Therefore one does not expect that $\Gamma_{\rm inh}$ would continue
to increase strongly for that last point, as the Ioffe-Regel
saturation must be felt at that frequency.
Fitting the first four points 
we find $\alpha$ = 5.7 $\pm$ 1.0 with $\chi^2$ = 0.28.
The best fit, shown by the full line, goes clearly above the last point
by an amount which is consistent with an $\Omega_{\rm co}$ that would
be located around 8.5 meV.

\begin{figure}
\centering\includegraphics[width=12cm]{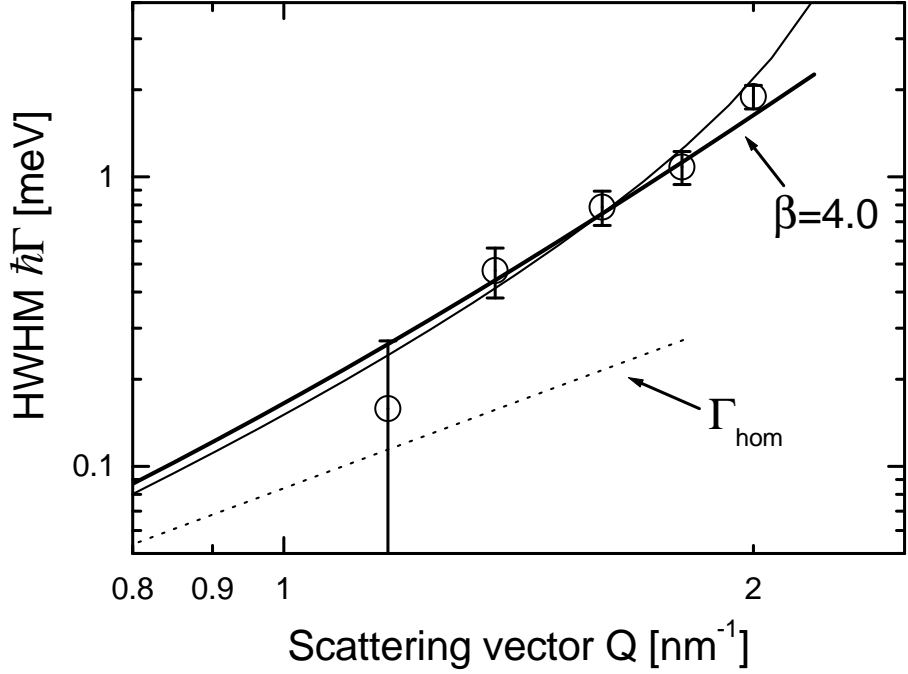}
\caption{The same half-widths presented in function of $Q$.
An estimated homogeneous broadening is shown by the dotted line labelled
$\Gamma_{\rm hom}$.
The heavy solid line is an adjustement of the first four points
to $BQ^{\beta}$.
The lighter solid line is an adjustment of the same points to equation (1).}
\end{figure}

The above evidence shows undisputably that there is a $\Gamma_{\rm inh}$
that grows rapidly between $\Omega \sim$ 4 and $\Omega \sim$ 8 meV
and that it tends to saturate above 8 meV.
It is of interest to also draw the same information in function of $Q$.
This is shown in figure 6.
Here we have traced $\Gamma_{\rm hom} \propto Q^2$.
The fit of the first four points to $\Gamma_{\rm inh}$ = B $\Omega ^{\beta}$
gives now $\beta =  4.0 \pm 1.2$ with $\chi^2 =  0.62$.
This difference with the power $\alpha$ above results from
the dispersion, $\Omega(q)$.
One must therefore be quite careful in discussing quantitatively the
approach to the crossover.
In fact, a more correct expression for disorder induced broadening
by Rayleigh scattering in a dispersive isotropic medium,
instead of $\Gamma_{\rm inh} \propto \Omega ^4$, is rather \cite{Tag03}

\begin{equation}
\Gamma_{\rm inh} = \Delta \; \Omega \; (aq)^2 \;
\left( \frac{\Omega a}{v_{\rm g}} \right) \;  \ .
\end{equation}

\noindent
Here, $\Delta$ is a dimensionless coefficient that characterizes
the disorder and $a$ is a microscopic dimension.
Equation (1) can also be adjusted to the first four data points in figure 6,
using the phase velocity $v_{\phi} = \Omega/q$ and the group velocity
$v_{\rm g}$ known from our measurement.
This gives the lighter solid line.
With $a = 1 / q_{\rm co}$ = 0.5 nm, we find $\Delta$ = 0.117 $\pm$ 0.012.
The $\chi^2$ equals 0.66 which is very satisfactory for this
one-parameter adjustment.
Knowing $\Delta$, and using $\ell^{-1} = D \Omega^4 (\hbar/k_{\rm B})^4$
as in \cite{Ray89}, one can extract

\begin{equation}
D = 2 \Delta \; a^3 \; k_{\rm B}^4 / v_{\phi}^2 v_{\rm g}^2 \hbar^4 \; \; .
\end{equation}

\noindent
This gives $D$ = 4 m$^{-1}$K$^{-4}$.
This value should be compared to $D$ = 100 m$^{-1}$K$^{-4}$ obtained in
\cite{Ray89} for $v$-SiO$_2$.
The lower value of $D$ in $d$-SiO$_2$ is of course consistent with the
higher value of $\Omega_{\rm co}$.
Following the analysis of \cite{Ray89}, this value is sufficient to
account for the $\kappa (T)$ plateau of $d$-SiO$_2$.
However, whether the size of $D$ can be explained as in \cite{Ray89}
requires additional information
about $d$-SiO$_2$ which does not seem available so far.\\

\section{Conclusions}

The main conclusion of this study is that a crossover to strong
scattering is definitely observed in $d$-SiO$_2$.
It is marked by a very rapid growth of the Brillouin linewidth
as the Brillouin frequency approaches $\Omega_{\rm co}$.
This result disagrees with previous claims that sound {\em propagates}
up to very high frequencies in silica \cite{Ben96,Mas97,Pil00}, but it
is in line with more recent observations on other strong glasses
\cite{Mat01a,Mat01b}.
For $v$-SiO$_2$, it also agrees with the result of a high-quality
simulation \cite{Tar00}.
More generally, our experience shows that IXS alone cannot prove
or disprove the existence of a $\Omega_{\rm co}$ if the latter is located
too low in frequency.
In this respect, other IXS-based claims \cite{Mas96,Mas98,Fio99,Ruo99,Mas00}
that sound propagates up to very high frequencies in glasses
that are otherwise known to show a well
defined $\kappa (T)$ plateau should be taken with great caution.
Only when the plateau is very weak and the BP almost absent, like
in calcium potassium nitrate CKN \cite{Mat01b},
may it be reasonable to think that acoustic waves propagate above
the BP.

A second strong conclusion is that the position of $\Omega_{\rm co}$
nearly coincides with the center of the boson peak, $\Omega_{\rm BP}$.
Our observations are in line with
a boson peak that consist of additional modes corresponding
to nearly rigid tetrahedra rotations.
These modes are strongly inhomogeneously broadened and their eigenvectors
should be highly disordered.

A third conclusion, not as strong as the previous ones because only
longitudinal excitations have been measured, is that the size
of the increase of $\Gamma_{\rm inh}$ with $\Omega$ or $q$ is in principle
sufficient to account for the thermal conductivity plateau.

The remaining question is that of the exact origin of this crossover.
Is it due to Rayleigh scattering, as advocated by many authors including
ourselves in earlier times \cite{For96,For97,Rat99}, is it rather
caused by hybridization with the Boson peak, or should it be
described by the soft-potential model \cite{Kar85,Par94,Ram98,Par01}?
In this respect one must remark that the random forces invoked in
\cite{Ray89} to explain the crossover in terms of Rayleigh scattering are
precisely the ${\rm Si-O-Si}$
bending forces which are the dominant restoring
forces for tetrahedra rigid rotations.
Presumably, the soft-potential and two-level systems are also
related to nearly rigid tetrahedra motions.
From the discussion in the previous section it seems that the
shape of the $Q$-dependence of $\Gamma_{\rm inh}$ will not
be able to settle this central issue.
However, the important forces have been quite well identified.
A critical theoretical analysis of the situation will
be necessary to make further progress.

\ack

The authors thank Dr. M. Arai for the excellent sample of $d$-SiO$_2$
without which these investigations would not have been possible.
They also thank the staff of ESRF, and in particular Dr. G. Monaco and
Dr. C. Masciovecchio for their expert supervision of the various IXS
experiments.
Prof. K. Inoue and Dr. Y. Yamanaka are thanked for their invaluable
collaboration in the hyper-Raman measurements performed in Sapporo.
Prof. A. K. Tagantsev is thanked for the derivation of the Rayleigh
scattering expression for a dispersive isotropic medium.

\end{document}